\documentclass[prd,showpacs,showkeys,nofootinbib,twocolumn]{revtex4-1}

\usepackage{amsmath,amsfonts,amssymb}
\usepackage[utf8]{inputenc}

\begin{document}

\title{Kerr Analogue of Kinnersley's Field of an Arbitrarily Accelerating Point Mass}

\author{Peter A. Hogan}
\email{peter.hogan@ucd.ie}
\affiliation{School of Physics, University College Dublin, Belfield, Dublin 4, Ireland} 

\author{Dirk Puetzfeld}
\email{dirk.puetzfeld@zarm.uni-bremen.de}
\homepage{http://puetzfeld.org}
\affiliation{University of Bremen, Center of Applied Space Technology and Microgravity (ZARM), 28359 Bremen, Germany} 

\date{ \today}

\begin{abstract}
We construct the field of an arbitrarily accelerating and rotating point mass which specializes to the Kerr solution when the acceleration vanishes and specializes to Kinnersley's arbitrarily accelerating point mass when the rotation vanishes.  
\end{abstract}


\maketitle


\section{Introduction}\label{sec_introduction}

Although Kinnersley's \cite{Kinnersley:1969} field of an arbitrarily accelerating point mass has been known for some time, and extensions of it to include charge have been given by Bonnor and Vaidya \cite{Bonnor:Vaidya:1972}, there does not appear to be an extension that specializes to the Kerr solution \cite{Kerr:1963} when the acceleration vanishes. There exist ``radiating Kerr metrics" due to Vaidya and Patel \cite{Vaidya:Patel:1973} and, more generally, due to Herlt \cite{Herlt:1980} (see \cite{Stephani:etal:2006} for a full description of what is available) but none appear to supply Kinnersley's field with spin included. The original solution by Kinnersley was also extensively studied in subsequent works on particles which accelerate by photon emission, so-called photon rockets, in the context of General Relativity, see for example \cite{Bonnor:1994,Damour:1995,Cornish:etal:1996,Bonnor:1996,Cornish:2000,Podolsky:2008,Podolsky:2011}.  

The geometrical construction of Kinnersley's field is described in detail in \cite{Stephani:etal:2006}. We require a modification of this construction to include spin. Since we are primarily interested in introducing acceleration of the source we will leave the mass and angular momentum per unit mass constant throughout. The construction here of the Kerr analogue of Kinnersley's model is a spin--off from the study of the equations of motion of a small, slowly rotating Kerr particle moving in an external gravitational field \cite{Hogan:2015}. In particular the geometrical construction in the present work is motivated by a construction of the Kerr solution with three components of angular momentum in \cite{Hogan:1977}. 

The structure of the paper is as follows: In section \ref{sec_kinnersley} we give a description of the construction of Kinnersley's field of an arbitrarily accelerating mass point as important background for our construction and to establish our notations and sign conventions. In section \ref{sec_kinnersley_kerr} we describe the geometrical construction of the Kerr analogue of Kinnersley's field and properties of the new model are given in section \ref{sec_kinnersley_kerr_properties}. Our results are concisely summarized in section \ref{sec_conclusions}.

\section{Kinnersely's Field}\label{sec_kinnersley}

We start with the Minkowskian line element in rectangular Cartesian coordinates and time $X^i=(X, Y, Z, T)$ for $i=1, 2, 3, 4$ (and we will use units for which the speed of light in a vacuum $c=1$ and the gravitational constant $G=1$):
\begin{equation}\label{kp1}
ds_0^2=-(dX)^2-(dY)^2-(dZ)^2+(dT)^2=\eta_{ij}\, dX^i\,dX^j\ .
\end{equation}
The Schwarzschild solution \cite{Schwarzschild:1916} of Einstein's vacuum field equations is given, in Kerr--Schild form \cite{Kerr:Schild:1965}, by the metric tensor
\begin{equation}\label{kp2}
g_{ij}=\eta_{ij}-\frac{2\,m}{R}\,k_i\,k_j\ ,
\end{equation}
with $k_i\,dX^i=dT-dR$, and $R=(X^2+Y^2+Z^2)^{1/2}$. Here $m$ $(={\rm constant})$ is the mass of the spherically symmetric source. With $k^i=\eta^{ij}\,k_j$ (and $\eta^{ij}$ defined by $\eta^{ij}\,\eta_{jk}=\delta^i_k$) we have $k^i\,k_i=0$. Also the inverse of the metric tensor (\ref{kp2}) (denoted $g^{ij}$ with $g^{ij}\,g_{jk}=\delta^i_k$) is given by
\begin{equation}\label{kp3}
g^{ij}=\eta^{ij}+\frac{2\,m}{R}\,k^i\,k^j\ ,
\end{equation}
and thus $k^i=\eta^{ij}\,k_j=g^{ij}\,k_j$ so that $k^i$ is a null vector field in the flat space--time with metric tensor $\eta_{ij}$ and in the curved space--time with metric tensor $g_{ij}$. We see that in the flat space--time with line element (\ref{kp1}) $R=0\ \Leftrightarrow\ X=Y=Z=0$ is a time--like geodesic (the $T$--axis). Effectively Kinnersley's \cite{Kinnersley:1969} construction replaces this geodesic with an arbitrary time--like world line. To achieve this let $X^i=w^i(u)$ be an arbitrary time--like world line in Minkowskian space--time with $u$ arc length or proper--time along it. Then $v^i(u)=dw^i/du$ is the unit time--like tangent to this world line satisfying $\eta_{ij}\,v^i\,v^j=v_j\,v^j=+1$. Thus $v^i$ is the 4--velocity of the particle with world line $X^i=w^i(u)$. The corresponding 4--acceleration is $a^i(u)=dv^i/du$ and satisfies $a_i\,v^i=0$ since $v_i\,v^i=1$. The position 4--vector of any point of Minkowskian space--time relative to the world line $X^i=w^i(u)$ may be written in the form
\begin{equation}\label{kp4}
X^i=w^i(u)+r\,k^i\ ,
\end{equation}
for $0\leq r<+\infty$ with $k^i\,k_i=0$ and $k^i\,v_i=+1$. Thus $k^i$ is a normalized future pointing null vector field defined along the world line $X^i=w^i(u)$ and so $k^i$ is tangent to the future null--cones with vertices on $X^i=w^i(u)$. The \emph{direction} of $k^i$ is parametrized by two real parameters $\xi$, $\eta$ (say) so that we can write
\begin{equation}\label{kp5}
k^i=P_0^{-1}\left (-\xi, -\eta, -1+\frac{1}{4}(\xi^2+\eta^2), 1+\frac{1}{4}(\xi^2+\eta^2)\right )\ ,
\end{equation}
for $-\infty<\xi, \eta<+\infty$ and with $P_0$ determined by the normalization $k^i\,v_i=1$ of $k^i$ to read
\begin{eqnarray}
P_0&=&v^1(u)\,\xi+v^2(u)\,\eta+v^3(u)\left (1-\frac{1}{4}(\xi^2+\eta^2)\right ) \nonumber \\
&&+v^4(u)\left (1+\frac{1}{4}(\xi^2+\eta^2)\right )\ .\label{kp6}
\end{eqnarray}
We note that
\begin{eqnarray}
h_0&\equiv&a_i\,k^i=P_0^{-1}\Biggl \{a^1(u)\,\xi+a^2(u)\,\eta\nonumber\\
&+&a^3(u)\left (1-\frac{1}{4}(\xi^2+\eta^2)\right ) +a^4(u)\left (1+\frac{1}{4}(\xi^2+\eta^2)\right )\Biggr \}\nonumber\\
&=&\frac{\partial}{\partial u}(\log P_0)\ ,\label{kp7}\end{eqnarray}
and thus the propagation law for $k^i$ along $X^i=w^i(u)$ is
\begin{equation}\label{kp8}
\frac{\partial k^i}{\partial u}=-h_0\,k^i\ .
\end{equation}
From (\ref{kp4}) we have
\begin{equation}\label{kp9}
dX^i=(v^i-r\,h_0\,k^i)du+k^i\,dr+r\,\frac{\partial k^i}{\partial\xi}d\xi+r\,\frac{\partial k^i}{\partial \eta}d\eta\ .
\end{equation}
Hence, in particular, 
\begin{equation}\label{kp10}
k_i\,dX^i=du\ \ \ \Leftrightarrow\ \ \ k_i=u_{,i}\ ,
\end{equation}
with the comma denoting partial differentiation with respect to $X^i$. Noting that
\begin{equation}\label{kp11}
\frac{\partial k^i}{\partial\xi}\frac{\partial k_i}{\partial\xi}=-P_0^{-2}=\frac{\partial k^i}{\partial\eta}\frac{\partial k_i}{\partial\eta}\ \ \ {\rm and}\ \ \frac{\partial k^i}{\partial\xi}\frac{\partial k_i}{\partial\eta}=0\ ,
\end{equation}
we have
\begin{eqnarray}
ds_0^2=\eta_{ij}\,dX^i\,dX^j&=&-r^2P_0^{-2}(d\xi^2+d\eta^2)+2\,du\,dr\nonumber \\
&&+(1-2\,h_0\,r)du^2\ .\label{kp12}
\end{eqnarray}
If we now generalize the metric tensor (\ref{kp3}) to provide the line element, in coordinates $x^i=(\xi, \eta, r, u)$,
\begin{eqnarray}
ds^2=g_{ij}\,dx^i\,dx^j&=&-r^2P_0^{-2}(d\xi^2+d\eta^2)+2\,du\,dr\nonumber \\
&&+\left (1-2\,h_0\,r-\frac{2\,m}{r}\right )du^2\ ,\label{kp13}
\end{eqnarray}
we arrive at Kinnersley's line element which reduces to the Eddington--Finkelstein form of the Schwarzschild line element when $a^i=0$ (i.e.\ when the world line $X^i=w^i(u)$ in Minkowskian space--time is a time--like geodesic). For this special case we may take $v^i=\delta^i_4$ and so $P_0=1+\frac{1}{4}(\xi^2+\eta^2)=p_0$ (say). Then (\ref{kp13}) becomes
the Schwarzschild line element
\begin{equation}\label{kp14}
ds^2=-r^2p_0^{-2}(d\xi^2+d\eta^2)+2\,du\,dr+\left (1-\frac{2\,m}{r}\right )du^2\ .
\end{equation}
Here $\xi$, $\eta$ are stereographic coordinates on the unit 2--sphere, related to the polar angles $\theta$, $\phi$ for $0\leq\theta\leq\pi, 0\leq\phi<2\,\pi$ via $\eta/\xi=\tan\phi$ and $\cos\theta=(4-\xi^2-\eta^2)/(4+\xi^2+\eta^2)$. We can write (\ref{kp13}) in terms of basis 1--forms $\vartheta^{(a)}$ with $a=1, 2, 3, 4$ as
\begin{equation}\label{kp15}
ds^2=-(\vartheta^{(1)})^2-(\vartheta^{(2)})^2+2\,\vartheta^{(3)}\,\vartheta^{(4)}=g_{(a)(b)}\,\vartheta^{(a)}\,\vartheta^{(b)}\ ,
\end{equation}
with
\begin{eqnarray}
\vartheta^{(1)}&=&r\,P_0^{-1}d\xi=-\vartheta_{(1)}\ ,\nonumber \\ 
\vartheta^{(2)}&=&r\,P_0^{-1}d\eta=-\vartheta_{(2)}\ ,\nonumber\\
\vartheta^{(3)}&=&du=\vartheta_{(4)}\ , \nonumber \\
\vartheta^{(4)}&=&dr+\left (\frac{1}{2}-h_0\,r-\frac{m}{r}\right )du=\vartheta_{(3)}\ ,\label{kp16} 
\end{eqnarray}
and $\vartheta_{(a)}=g_{(a)(b)}\,\vartheta^{(b)}$. The 1--forms define a half null tetrad and the components $R_{(a)(b)}$ of the Ricci tensor on this tetrad vanish with the exception of 
\begin{equation}\label{kp17}
R_{(3)(3)}=-\frac{6\,m\,h_0}{r^2}\ .
\end{equation}
Hence if $R_{ij}$ are the components of the Ricci tensor in coordinates $x^i=(\xi, \eta, r, u)$ then we can write
\begin{equation}\label{kp18}
R_{ij}=-\frac{6\,m\,h_0}{r^2}\,n_i\,n_j\ \ \ {\rm with}\ \ \ n_i\,dx^i=du\ .
\end{equation}
We note that $g^{ij}\,n_i\,n_j=0$.

In preparation for consideration of the axially symmetric Kerr case it is useful to specialize the Kinnersley model to the case for which the world line $X^i=w^i(u)$ is the history of a particle performing rectilinear motion. For this we take the 4--velocity $v^i$ to be restricted by requiring $v^1(u)=0=v^2(u)$. Then writing $\lambda=v^4+v^3$ and $\mu=v^4-v^3$ we have $\lambda\,\mu=1$ and, with a dot denoting differentiation with respect to $u$, we see that if 
\begin{equation}\label{kp19}
\frac{\dot\lambda}{\lambda}=A(u)\ \ \ {\rm then}\ \ \ A^2=-a_i\,a^i=(a^3)^2-(a^4)^2\ .
\end{equation}
Now $P_0$ and $h_0$ in (\ref{kp6}) and (\ref{kp7}) read
\begin{equation}\label{kp19'}
P_0=\lambda^{-1}\left (\lambda^2+\frac{1}{4}(\xi^2+\eta^2)\right )\ ,
\end{equation}
and
\begin{equation}\label{kp19''}
h_0=A(u)\left (\frac{\lambda^2-\frac{1}{4}(\xi^2+\eta^2)}{\lambda^2+\frac{1}{4}(\xi^2+\eta^2)}\right )\ .
\end{equation}
We introduce polar coordinates $\theta$, $\phi$ by writing
\begin{eqnarray}
\xi&=&2\,\lambda\,\left (\frac{1-\cos\theta}{1+\cos\theta}\right )^{1/2}\cos\phi\ ,\nonumber \\
\eta&=&2\,\lambda\,\left (\frac{1-\cos\theta}{1+\cos\theta}\right )^{1/2}\sin\phi\ ,\label{kp20}
\end{eqnarray}
and thus
\begin{equation}\label{kp21}
\frac{\eta}{\xi}=\tan\phi\ \ \ {\rm and}\ \ \ \frac{\lambda^2-\frac{1}{4}(\xi^2+\eta^2)}{\lambda^2+\frac{1}{4}(\xi^2+\eta^2)}=\cos\theta\ .
\end{equation}
This results in $h_0=A(u)\,\cos\theta$,
\begin{eqnarray}
P_0^{-1}d\xi&=&-\sin\theta\,\sin\phi\,d\phi+(d\theta+A(u)\,\sin\theta\,du)\cos\phi\ ,\nonumber\\
\label{kp22}\\
P_0^{-1}d\eta&=&\sin\theta\,d\phi\,\cos\phi+(d\theta+A(u)\,\sin\theta\,du)\sin\phi\ ,\nonumber\\
\label{kp23}
\end{eqnarray}
and thus the line element (\ref{kp13}) takes the form
\begin{eqnarray}
ds^2&=&-r^2\{(d\theta+A(u)\,\sin\theta\,du)^2+\sin^2\theta\,d\phi^2\}+2\,du\,dr\nonumber\\
&&+\left (1-2\,A(u)\,r\,\cos\theta-\frac{2\,m}{r}\right )du^2\ ,\nonumber\\
&=&\left (\underset{(0)}{g_{ij}}-\frac{2\,m}{r}\,n_i\,n_j\right )dx^i\,dx^j\ ,\nonumber\\
&=&g_{ij}\,dx^i\,dx^j\ .\label{kp24}
\end{eqnarray}
Here the Minkowskian metric tensor components in the curvilinear coordinates $x^i=(\theta, \phi, r, u)$ are denoted $\underset{(0)}{g_{ij}}$. The Minkowskian metric tensor components in the rectangular Cartesian coordinates and time $X^i=(X, Y, Z, T)$ are denoted $\eta_{ij}$ and thus $\underset{(0)}{g_{ij}}\,dx^i\,dx^j=\eta_{ij}\,dX^i\,dX^j$. We note that $g_{ij}=\underset{(0)}{g_{ij}}-(2\,m/r)n_i\,n_j$ has Kerr--Schild form. Also $n^i=\underset{(0)}{g^{ij}}\,n_j=g^{ij}\,n_j=\delta^i_3$. The Kerr--Schild form of the metric tensor has the important algebraic property of the equality of determinants:
\begin{equation}\label{kp24'}
g={\rm det}(g_{ij})={\rm det}(\underset{(0)}{g_{ij}})=\underset{(0)}{g}=-r^4\sin^2\theta\ .
\end{equation}
Hence, if covariant differentiation with respect to the Riemannian connection calculated with the metric tensor $g_{ij}$ is denoted by a stroke, we have 
\begin{equation}\label{kp24''}
n^i{}_{|i}=\frac{1}{2\,g}\,\frac{\partial g}{\partial r}=\frac{1}{2\,\underset{(0)}{g}}\,
\frac{\partial}{\partial r}(\underset{(0)}{g})=\frac{2}{r}\ ,\end{equation}
and so, for future reference, we see that the metric tensor above has the form
\begin{equation}\label{kp24'''}
g_{ij}=\underset{(0)}{g_{ij}}-m\,n^k{}_{|k}\,n_i\,n_j\ .
\end{equation}
We note that the optical scalar describing the expansion of the congruence of null geodesic integral curves of the vector field $n^i$ is $(1/2)n^i{}_{|i}$.

\section{Kerr analogue of Kinnersley's field}\label{sec_kinnersley_kerr}

To introduce rotation or spin into the Kinnersley field we proceed by modifying (\ref{kp4}) to read
\begin{equation}\label{kp25'}
X^i=w^i(u)+r\,k^i+U^i\  ,
\end{equation}
with 
\begin{equation}\label{kp25}
   U^i(\xi, \eta, u)=P_0^2\left (\frac{\partial k^i}{\partial\xi}\frac{\partial F}{\partial \eta}-\frac{\partial k^i}{\partial\eta}\frac{\partial F}{\partial \xi}\right )\ .
\end{equation}
Here $k^i$ and $P_0$ are given by (\ref{kp5}) and (\ref{kp6}) and 
\begin{equation}\label{kp26}
F=s^i(u)\,k_i\ \ {\rm with}\ \ s^i\,v_i=0\ \ \ {\rm and}\ \ \ \frac{ds^i}{du}=-(a_j\,s^j)\,v^i\ .
\end{equation}
The space--like vector $s^i(u)$ is here defined along the world line $X^i=w^i(u)$ by Fermi transport. We note the useful formulas:
\begin{eqnarray}
\frac{\partial U^i}{\partial\xi}&=&(v^i-k^i)\frac{\partial F}{\partial\eta}+\frac{\partial k^i}{\partial\eta}F\ ,\label{kp27}\\
\frac{\partial U^i}{\partial\eta}&=&-(v^i-k^i)\frac{\partial F}{\partial\xi}-\frac{\partial k^i}{\partial\xi}F\ ,\label{kp28}\\
\frac{\partial U^i}{\partial u}&=&P_0^2\left (\frac{\partial h_0}{\partial \eta}\frac{\partial F}{\partial\xi}-\frac{\partial h_0}{\partial \xi}\frac{\partial F}{\partial\eta}\right )k^i\nonumber \\
&&+F\,P_0^2\left (\frac{\partial h_0}{\partial\xi}\frac{\partial k^i}{\partial\eta}-\frac{\partial h_0}{\partial\eta}\frac{\partial k^i}{\partial\xi}\right )\ .\label{kp29}\end{eqnarray}
We now, for simplicity, specialize the world line $X^i=w^i(u)$ by requiring $v^1(u)=0=v^2(u)$ as we did at the end of the previous section. Then one can solve the propagation law (\ref{kp26}) for $s^i$ along $X^i=w^i(u)$ with
\begin{equation}\label{kp30}
s^i(u)=\left (0, 0, \frac{1}{2}\,\underset{(0)}{S}(\lambda+\lambda^{-1}), \frac{1}{2}\,\underset{(0)}{S}(\lambda-\lambda^{-1})\right )\ ,
\end{equation}
with $\lambda(u)=v^4+v^3$ as before and $\underset{(0)}{S}={\rm constant}$. We then find that 
\begin{equation}\label{kp31}
F=s^i\,k_i=\underset{(0)}{S}\left (\frac{\lambda^2-\frac{1}{4}(\xi^2+\eta^2)}{\lambda^2+\frac{1}{4}(\xi^2+\eta^2)}\right )\ ,\end{equation}
and $P_0$, $h_0$ are given by (\ref{kp19'}). Now introducing the polar angles $\theta$, $\phi$ via (\ref{kp20}) we have 
$h_0=A(u)\,\cos\theta$ as before and $F=\underset{(0)}{S}\,\cos\theta$ and we obtain from (\ref{kp25})
\begin{eqnarray}
dX^i&=&(du+\underset{(0)}{S}\,\sin^2\theta\,d\phi)v^i\nonumber\\
&&+(dr-r\,A\,\cos\theta\,du-\underset{(0)}{S}\,\sin^2\theta\,d\phi)\,k^i\nonumber\\
&&+(\lambda^{(1)}\cos\phi-\lambda^{(2)}\sin\phi)P_0\frac{\partial k^i}{\partial\xi}\nonumber\\
&&+(\lambda^{(1)}\sin\phi+\lambda^{(2)}\cos\phi)P_0\frac{\partial k^i}{\partial\eta}\ , \label{kp32}
\end{eqnarray}
with the 1--forms $\lambda^{(1)}$, $\lambda^{(2)}$ given by
\begin{eqnarray}
\lambda^{(1)}&=&r\,(d\theta+A(u)\,\sin\theta\,du)-\underset{(0)}{S}\,\sin\theta\cos\theta\,d\phi\ ,\label{kp33}\\
\lambda^{(2)}&=&r\,\sin\theta\,d\phi+\underset{(0)}{S}\,\cos\theta\,d\theta\ .\label{kp34}
\end{eqnarray}
We note that
\begin{equation}\label{kp35}
k_i\,dX^i=du+\underset{(0)}{S}\,\sin^2\theta\,d\phi=n_i\,dx^i \ ,
\end{equation}
defining $n_i$ in coordinates $x^i=(\theta, \phi, r, u)$ and, using the Minkowskian scalar products (\ref{kp11}), 
\begin{eqnarray}
\eta_{ij}\,dX^i\,dX^j&=&-(\lambda^{(1)})^2-(\lambda^{(2)})^2+2\,(du+\underset{(0)}{S}\,\sin^2\theta) \nonumber \\
&& \times \Biggl\{dr-r\,A\,\cos\theta\,du-\underset{(0)}{S}\,\sin^2\theta\,d\phi\nonumber\\
&& +\frac{1}{2}(du+\underset{(0)}{S}\,\sin^2\theta\,d\phi)\Biggr\}\ ,\nonumber\\
&=&\underset{(0)}{g_{ij}}\,dx^i\,dx^j\ .\label{kp36}
\end{eqnarray}
When $\underset{(0)}{S}=0$ this line element coincides with (\ref{kp24}) when $m=0$. When $A=0$ in (\ref{kp36}) the line element is the Minkowskian \emph{background} for the Kerr solution with angular momentum per unit mass $\underset{(0)}{S}$. The constant $\underset{(0)}{S}$ is thus playing the role of the constant $a$ in the standard form of the Kerr solution. We are using $\underset{(0)}{S}$ to denote the angular momentum per unit mass or spin rather than the more familiar $a$ to avoid any confusion with the acceleration. We now form the Kerr--Schild metric tensor (\ref{kp24'''}) in this case with $n_i$ given by (\ref{kp35}). Once again we have $n^i=\delta^i_3$ and using (\ref{kp24''}) we find that 
\begin{equation}\label{kp36'}
n^i{}_{|i}=\frac{2\,r+A\,\underset{(0)}{S}^2\sin^2\theta\,\cos\theta}{r^2+\underset{(0)}{S}^2\cos^2\theta+r\,A\,
\underset{(0)}{S}^2\sin^2\theta\,\cos\theta}\ .
\end{equation}
Substituting (\ref{kp36}) and (\ref{kp36'}) into (\ref{kp24'''}) we arrive at the Kerr analogue of Kinnersley's model which can be written
\begin{eqnarray}\label{kp38}
ds^2&=&g_{ij}\,dx^i\,dx^j=-(\lambda^{(1)})^2-(\lambda^{(2)})^2+2\,\lambda^{(3)}\,\lambda^{(4)}\nonumber \\
&=&g_{(a)(b)}\,\lambda^{(a)}\,\lambda^{(b)}\ ,
\end{eqnarray}
with $\lambda^{(1)}$, $\lambda^{(2)}$ given by (\ref{kp33}) and (\ref{kp34}) and
\begin{eqnarray}
\lambda^{(3)}&=&du+\underset{(0)}{S}\,\sin^2\theta\,d\phi\ ,\label{kp39}\\
\lambda^{(4)}&=&dr-r\,A(u)\,\cos\theta\,du-\underset{(0)}{S}\,\sin^2\theta\,d\phi \nonumber \\
&&+\left (\frac{1}{2}-\frac{m\,\varphi'}{2\,\varphi}\right )\lambda^{(3)}\ ,\nonumber\\\label{kp40}
\end{eqnarray}
with
\begin{eqnarray}\label{kp40'}
\varphi=r^2+\underset{(0)}{S}^2\cos^2\theta+r\,A\,\underset{(0)}{S}^2\sin^2\theta\,\cos\theta\ ,
\end{eqnarray}
and $\varphi'=\partial\varphi/\partial r$. We emphasize that $A(u)$ is an arbitrary function of its argument while $m$ and $\underset{(0)}{S}$ are constants.

Writing the basis 1--forms $\lambda^{(a)}$ given by (\ref{kp33}), (\ref{kp34}), (\ref{kp39}) and (\ref{kp40}) in the form
\begin{eqnarray}
\lambda^{(1)}&=&f_i\,dx^i\ ,\quad \quad \lambda^{(2)}=e_i\,dx^i\ ,\ \nonumber \\
\lambda^{(3)}&=&n_i\,dx^i\ ,\quad \quad \lambda^{(4)}=l_i\,dx^i\ ,\label{kp46}
\end{eqnarray}
with $x^i=(\theta, \phi, r, u)$, all scalar products among the basis vectors $f^i, e^i, n^i, l^i$ vanish except 
$f_i\,f^i=e_i\,e^i=-n_i\,l^i=-1$. An exact calculation of the Ricci tensor components $R_{ij}$ in coordinates $x^i$ results in 
\begin{eqnarray}
R_{ij}&=&R_{(1)(1)}\,f_i\,f_j+R_{(2)(2)}\,e_i\,e_j+R_{(1)(2)}\,(f_i\,e_j+f_j\,e_i)\nonumber\\
&&+R_{(1)(3)}\,(f_i\,n_j+f_j\,n_i)+R_{(2)(3)}\,(e_i\,n_j+e_j\,n_i)\nonumber \\
&&+R_{(3)(4)}\,(n_i\,l_j+n_j\,l_i)+R_{(3)(3)}\,n_i\,n_j\ ,\label{kp47}
\end{eqnarray}
with
\begin{eqnarray}
R_{(1)(1)}&=&m\,A^2\underset{(0)}{S}^4\sin^4\theta\,\cos^2\theta\,\varphi^{-3}\varphi'+m\,A\,\underset{(0)}{S}^2\nonumber\\
&&\times\sin^2\theta\cos\theta\,\varphi^{-3}(r^2-\underset{(0)}{S}^2\cos^2\theta)\ ,\label{kp48}\\
R_{(2)(2)}&=&-m\,A\,\underset{(0)}{S}^2\sin^2\theta\,\cos\theta\,\varphi^{-3}(r^2-\underset{(0)}{S}^2\cos^2\theta)\ ,\nonumber\\ \label{kp49}\\
R_{(1)(2)}&=&-m\,A\,\underset{(0)}{S}^3\sin^2\theta\,\cos^2\theta\,\varphi^{-3}\varphi'\ ,\label{kp50}\\
R_{(3)(4)}&=&\frac{1}{2}\,m\,A^2\underset{(0)}{S}^4\sin^4\theta\,\cos^2\theta\,\varphi^{-3}\varphi'\ ,\label{kp51}
\end{eqnarray}
and $\varphi$ is given by (\ref{kp40'}). The remaining tetrad components of the Ricci tensor to be calculated are $R_{(1)(3)}, R_{(2)(3)}$ and $R_{(3)(3)}$. These latter components are considerably more complicated than (\ref{kp48})--(\ref{kp51}) and are only required asymptotically (for large $r$) for our purposes and are given in (\ref{kp20a})--(\ref{kp22a}) below. We note that 
\begin{equation}\label{kp52}
g^{ij}\,R_{ij}=-R_{(1)(1)}-R_{(2)(2)}+2\,R_{(3)(4)}=0\ .
\end{equation}

The generalized Kerr congruence (i.e.\ the Kerr congruence generalized to include the influence of the acceleration)  consists of the integral curves of the null vector field $n$ which is given, in coordinates $x^i=(\theta, \phi, r, u)$ by
\begin{equation}\label{kp5a}
n=n^i\,\frac{\partial}{\partial x^i}=\frac{\partial}{\partial r}\ \ \ {\rm and}\ \ \ n_i\,dx^i=\underset{(0)}{S}\,\sin^2\theta\,d\phi+du\ .
\end{equation}
This is a geodesic congruence since
\begin{equation}\label{kp6a}
n^i{}_{|j}\,n^j=\frac{1}{2}\,g^{ij}\,(2\,g_{j3,3}-g_{33,j})=0\ ,
\end{equation}
and $g_{3i}=(0, \underset{(0)}{S}\,\sin^2\theta, 0, 1)$. Thus the integral curves of the vector field $n$ are null geodesics with $r$ an affine parameter along them. If we define the complex null vector $m^i$ and its complex conjugate $\bar m^i$ by
\begin{equation}\label{kp7a}
m^i=\frac{1}{\sqrt{2}}(f^i+i\,e^i)\ \ \ {\rm and}\ \ \  \bar m^i=\frac{1}{\sqrt{2}}(f^i-i\,e^i)\ ,
\end{equation}
then the complex shear $\sigma$ of this null geodesic congruence is given by
\begin{equation}\label{kp8a}
\sigma=n_{i|j}\,m^i\,m^j\ \ \ \Rightarrow\ \ |\sigma|^2=\frac{1}{2}n_{(i|j)}\,n^{i|j}-\Theta^2\ ,
\end{equation}
where the round brackets denote symmetrization and
\begin{equation}\label{kp8a}
\Theta=\frac{1}{2}\,n^i{}_{|i}\ ,
\end{equation}
is the expansion scalar of the congruence. Also 
\begin{equation}\label{kp9a}
\rho=n_{i|j}\,m^i\,\bar m^j=-\Theta+i\omega\ ,
\end{equation}
with
\begin{equation}\label{kp10a}
\omega^2=\frac{1}{2}\,n_{[i|j]}\,n^{i|j}\ ,
\end{equation}
the twist of the congruence. The square brackets here denote antisymmetrization. For the particular case of the generalized Kerr congruence the complex shear $\sigma$ is in fact real and is given by
\begin{equation}\label{kp11a}
\sigma=-\frac{1}{2}\,\varphi^{-1}A\,\underset{(0)}{S}^2\sin^2\theta\,\cos\theta\ ,
\end{equation}
while the complex scalar (\ref{kp9a}) takes the form
\begin{equation}\label{kp12a}
\rho=-\frac{1}{2}\varphi^{-1}\varphi'+i\,\varphi^{-1}\underset{(0)}{S}\,\cos\theta\ .
\end{equation}
We already know that the tetrad component $R_{(4)(4)}=R_{ij}\,n^i\,n^j$ vanishes. This can now be verified using the propagation equation for $\rho$ along the congruence:
\begin{equation}\label{kp13a}
\frac{\partial\rho}{\partial r}=\rho^2+\sigma^2+\frac{1}{2}\,R_{(4)(4)}\ .
\end{equation}
We can write $R_{(1)(1)}$, $R_{(2)(2)}$, $R_{(1)(2)}$, $R_{(3)(4)}$ in terms of the geometrical variables $\sigma$, $\omega$ as follows: Using
\begin{equation}\label{kp13a'}
\sigma=\underset{(0)}{\sigma}\,\varphi^{-1}\ \ \ {\rm with}\ \  \underset{(0)}{\sigma}=-\frac{1}{2}\,A\,\underset{(0)}{S}^2\sin^2\theta\,\cos\theta\ ,
\end{equation}
and
\begin{equation}\label{kp13a''}
\omega=\underset{(0)}{\omega}\,\varphi^{-1}\ \ \ {\rm with}\ \  \underset{(0)}{\omega}=\underset{(0)}{S}\,\cos\theta\ ,
\end{equation}
we find from (\ref{kp48})--(\ref{kp51}):
\begin{eqnarray}
R_{(1)(1)}&=&-2\,m\,\underset{(0)}{\sigma}\,\varphi^{-2}+4\,m\,\underset{(0)}{\sigma}^2r\,\varphi^{-3}\nonumber\\
&&+4\,m\,\underset{(0)}{\sigma}(\underset{(0)}{\omega}^2-2\,\underset{(0)}{\sigma}^2)\varphi^{-3}\ ,\label{kp14a}\\
R_{(2)(2)}&=&2\,m\,\underset{(0)}{\sigma}\,\varphi^{-2}+4\,m\,\underset{(0)}{\sigma}^2r\,\varphi^{-3}-4\,m\,\underset{(0)}{\sigma}\,\underset{(0)}{\omega}^2\,\varphi^{-3}\ ,\nonumber\\\label{kp15a}\\
R_{(1)(2)}&=&4\,m\,\underset{(0)}{\sigma}\,\underset{(0)}{\omega}\,r\,\varphi^{-3}-4\,m\,\underset{(0)}{\sigma}^2\underset{(0)}{\omega}\,\varphi^{-3}\ ,\label{kp16a}\\
R_{(3)(4)}&=&4\,m\,\underset{(0)}{\sigma}^2r\,\varphi^{-3}-4\,m\,\underset{(0)}{\sigma}^3\varphi^{-3}\ .\label{kp17a}
\end{eqnarray}
Here the dependence of the tetrad components of the Ricci tensor on the radial coordinate $r$ is explicit, remembering that $\varphi$ is given by (\ref{kp40'}) and since $\underset{(0)}{\sigma}, \underset{(0)}{\omega}$ are independent of $r$. The Ricci tensor components $R_{ij}$, in coordinates $x^i=(\theta, \phi, r, u)$, have the exact algebraic form given in (\ref{kp47}). We have noted following (\ref{kp47}) that the Ricci scalar $R=g^{ij}\,R_{ij}$ vanishes exactly and so the Ricci tensor $R_{ij}$ and the Einstein tensor $G_{ij}=R_{ij}-\frac{1}{2}\,g_{ij}\,R$ coincide. We now also note that the basis vector fields $f^i$, $e^i$, $l^i$, $n^i$ are parallel transported along the future--directed null geodesic integral curves of the vector field $n^i$ and thus
\begin{equation}\label{kp18a}
f^i{}_{|j}\,n^j=0\ ,\ e^i{}_{|j}\,n^j=0\ ,\ l^i{}_{|j}\,n^j=0\ \ {\rm and}\ \ n^i{}_{|j}\,n^j=0\ .
\end{equation}
For large positive values of the affine parameter $r$ we see from (\ref{kp14a})--(\ref{kp17a}) that 
\begin{eqnarray}
R_{(1)(1)}&=&O\left (\frac{1}{r^4}\right ),\quad \quad R_{(2)(2)}=O\left (\frac{1}{r^4}\right ),\nonumber\\
R_{(1)(2)}&=&O\left (\frac{1}{r^5}\right ),\quad \quad R_{(3)(4)}=O\left (\frac{1}{r^5}\right )\ .\label{kp19a}
\end{eqnarray}
The remaining non--vanishing tetrad components of the Ricci tensor, namely $R_{(1)(3)}$, $R_{(2)(3)}$ and $R_{(3)(3)}$, 
have the following asymptotic forms for large $r$:
\begin{eqnarray}
R_{(1)(3)}&=&O\left (\frac{1}{r^4}\right )\ ,\label{kp20a}\\
R_{(2)(3)}&=&-\frac{6\,m\,A\,\underset{(0)}{S}\,\sin\theta\,\cos\theta}{r^3}+O\left (\frac{1}{r^4}\right )\ ,\label{kp21a}\\
R_{(3)(3)}&=&-\frac{6\,m\,A\,\cos\theta}{r^2}+\frac{1}{r^3}\Biggl\{-6\,m\,\dot A\,\underset{(0)}{S}^2\sin^2\theta\,\cos\theta\nonumber\\
&&-6\,m\,A^2\underset{(0)}{S}^2\sin^2\theta+48\,m\,A^2\underset{(0)}{S}^2\sin^2\theta\,\cos^2\theta\Biggr\}\nonumber\\
&&+O\left (\frac{1}{r^4}\right )\ .\label{kp22a}
\end{eqnarray}
Here, as always, $\dot A=dA/du$. Substituting (\ref{kp19a})--(\ref{kp22a}) into (\ref{kp47}) and given the energy--momentum--stress tensor of matter $T_{ij}$ via Einstein's equations $R_{ij}=-8\,\pi\,T_{ij}$ we find that we can write
\begin{equation}\label{kp23a}
T_{ij}=\underset{(1)}{T_{ij}}+\underset{(2)}{T_{ij}}+O\left (\frac{1}{r^4}\right )\ ,
\end{equation}
with
\begin{equation}\label{kp24a}
8\,\pi\,\underset{(1)}{T_{ij}}=\left (\frac{6\,m\,A\,\cos\theta}{r^2}-\frac{6\,m\,A^2\underset{(0)}{S}^2\sin^2\theta\,\cos^2\theta}{r^3}\right )n_i\,n_j,
\end{equation}
and
\begin{eqnarray}
&&8\,\pi\,\underset{(2)}{T_{ij}}=\frac{6\,m\,A\,\underset{(0)}{S}\,\sin\theta\,\cos\theta}{r^3}\Biggl (e_i\,n_j+e_j\,n_i\Biggr )\nonumber \\
&&+\frac{1}{r^3}\Biggl (6\,m\,\dot A\,\underset{(0)}{S}^2\sin^2\theta\,\cos\theta-42\,m\,A^2\underset{(0)}{S}^2\sin^2\theta\,\cos^2\theta\nonumber\\
&&+6\,m\,A^2\underset{(0)}{S}^2\sin^2\theta\Biggr )n_i\,n_j\ .\label{kp25a}
\end{eqnarray}
We have chosen the definitions of $\underset{(1)}{T_{ij}}$ and $\underset{(2)}{T_{ij}}$ here because now they each separately satisfy the approximate conservation equations
\begin{equation}\label{kp26a}
\underset{(1)}{T^{ij}}{}_{|j}=O\left (\frac{1}{r^5}\right )\ \quad {\rm and}\quad \underset{(2)}{T^{ij}}{}_{|j}=O\left (\frac{1}{r^5}\right )\ .
\end{equation}
The first of these is easily verified using $n^i\frac{\partial}{\partial x^i}=\frac{\partial}{\partial r}$ and
\begin{equation}\label{kp27a}
n^i{}_{|i}=\frac{2}{r}-\frac{A\,\underset{(0)}{S}^2\sin^2\theta\,\cos\theta}{r^2}+O\left (\frac{1}{r^3}\right )\ .
\end{equation}
The leading $r^{-2}$--term in the energy--momentum--stress tensor here does not involve the spin parameter of the Kerr solution and coincides with the Kinnersley energy--momentum--stress tensor. It is algebraically identical to Kinnersley's energy--momentum--stress tensor if, instead of using the affine parameter distance $r$, we use the parallax distance $r_P$ defined by \cite{Sachs:1962}
\begin{equation}\label{kp27aa}
r_P=\Theta^{-1}=\frac{r^2+r\,A\,\underset{(0)}{S}^2\sin^2\theta\,\cos\theta+\underset{(0)}{S}^2\cos^2\theta}{r+\frac{1}{2}A\,\underset{(0)}{S}^2\sin^2\theta\,\cos\theta}\ .
\end{equation}
Now for large values of $r$ we have 
\begin{equation}\label{kp28aa}
\frac{1}{r_P^2}=\frac{1}{r^2}-\frac{A\,\underset{(0)}{S}^2\sin^2\theta\,\cos\theta}{r^3}+O\left (\frac{1}{r^4}\right )\ ,
\end{equation}
and so
\begin{equation}\label{kp29aa}
8\,\pi\,\underset{(1)}{T_{ij}}=\frac{6\,m\,A\,\cos\theta}{r_P^2}\,n_i\,n_j\ .
\end{equation}
With the shear $\sigma$ given by (\ref{kp11a}) and the twist $\omega$ and expansion $\Theta$ of the integral curves of the vector field $n^i$ given via (\ref{kp9a}) and (\ref{kp12a}) we have an exact formula relating the affine parameter distance $r$ and the parallax distance $r_P$:
\begin{equation}\label{kp30aa}
r=\frac{r_P(1+\sigma\,r_P)}{1+(\omega\,r_P)^2-(\sigma\,r_P)^2}\ .
\end{equation}
This generalizes a formula derived by Sachs \cite[eq.(23)]{Sachs:1962} in the shear--free case.

We note that $\underset{(1)}{T^{ij}}$ corresponds to the two leading terms in the expansion for large $r$ of
\begin{equation}\label{kp28a}
8\,\pi\,\underset{(1)}{t^{ij}}=6\,m\,A\,\varphi^{-1}\cos\theta\,n^i\,n^j\ ,
\end{equation}
with $\varphi$ given by (\ref{kp40'}), which satisfies the exact conservation equation
\begin{equation}\label{kp29a}
\underset{(1)}{{t}^{ij}}{}_{|j}=0\ ,
\end{equation}
on account of (\ref{kp36'}). Verifying the second of (\ref{kp26a}) requires
\begin{eqnarray}
n^i{}_{|j}\,e^j&=&-\varphi^{-1}r\,A\,\underset{(0)}{S}\sin\theta\,\cos\theta\,n^i+\varphi^{-1}r\,e^i\nonumber\\
&&+\varphi^{-1}\underset{(0)}{S}\cos\theta\,f^i\ ,\label{kp30a}\\
e^i{}_{|i}&=&-\varphi^{-1}r\,A\,\underset{(0)}{S}\sin\theta\,\cos\theta-\varphi^{-1}\underset{(0)}{S}\cot\theta\,\cos\theta\ ,\nonumber\\
\label{kp31a}\end{eqnarray}
with $\varphi$ expanded in inverse powers of $r$.

\section{Properties of the Kerr analogue of Kinnersley's field}\label{sec_kinnersley_kerr_properties}

To identify a curvature singularity in the Kerr analogue of Kinnersley's field a convenient approach is to calculate the Kretschmann scalar. This is given in our case by 
\begin{eqnarray}
&&R_{ijkl}\,R^{ijkl}=\frac{2\,m^2}{\varphi^6}\Biggl\{24\,r^6-360\,r^4\underset{(0)}{S}^2\cos^2\theta\nonumber\\
&&+360\,r^2\underset{(0)}{S}^4\cos^4\theta-24\,\underset{(0)}{S}^6\cos^6\theta+72\,r^5A\,\underset{(0)}{S}^2\sin^2\theta\,\cos\theta \nonumber\\
&&+104\,r^4A^2\underset{(0)}{S}^4\sin^4\theta\,\cos^2\theta-720\,r^3A\,\underset{(0)}{S}^4\sin^2\theta\,\cos^3\theta\nonumber\\
&&+88\,r^3A^3\underset{(0)}{S}^6\sin^6\theta\,\cos^3\theta+50\,r^2A^4\underset{(0)}{S}^8\sin^8\theta\,\cos^4\theta\nonumber\\
&&-608\,r^2A^2\underset{(0)}{S}^6\sin^4\theta\,\cos^4\theta+360\,r\,A\,\underset{(0)}{S}^6\sin^2\theta\,\cos^5\theta\nonumber\\
&&-248\,r\,A^3\underset{(0)}{S}^8\sin^6\theta\,\cos^5\theta+18\,r\,A^5\underset{(0)}{S}^{10}\sin^{10}\theta\,\cos^5\theta\nonumber\\
&&+3\,A^6\underset{(0)}{S}^{12}\sin^{12}\theta\,\cos^6\theta-42\,A^4\underset{(0)}{S}^{10}\sin^8\theta\,\cos^6\theta\nonumber\\
&&+104\,A^2\underset{(0)}{S}^8\sin^4\theta\,\cos^6\theta\Biggr\}\ ,\label{kp48a}
\end{eqnarray}
with $\varphi$ given by (\ref{kp40'}). Thus the Kretschmann scalar is singular when $\varphi=0$ and this occurs, in particular, at $r=0$, $\theta=\pi/2$. When (\ref{kp25}) is specialized to the axially symmetric case we have
\begin{eqnarray}
X&=&r\,k^1-\underset{(0)}{S}\,k^2\ ,\label{kp49a}\\
Y&=&r\,k^2+\underset{(0)}{S}\,k^1\ ,\label{kp50a}\\
Z&=&w^3(u)+r\,k^3\ ,\label{kp51a}\\
T&=&w^4(u)+r\,k^4\ .\label{kp52a}
\end{eqnarray}
Since $(k^1)^2+(k^2)^2=\sin^2\theta$ we see that when $r=0$ and $\theta=\pi/2$ the singularity in the curvature occurs on the ring
\begin{equation}\label{kp53a}
X^2+Y^2=\underset{(0)}{S}^2\ ,
\end{equation}
which is accelerating in the positive $Z$--direction when $A(u)\neq0$. 
More generally, using $\underset{(0)}{\sigma}$ and $\underset{(0)}{\omega}$ given by (\ref{kp13a'}) and (\ref{kp13a''}), we can write 
$\varphi$ in (\ref{kp40'}) in the form
\begin{equation}\label{kp55a}
\varphi=r^2-2\,r\,\underset{(0)}{\sigma}+\underset{(0)}{\omega}^2\ .
\end{equation}
It therefore follows from (\ref{kp48a}) that curvature singularities occur when
\begin{equation}\label{kp56a}
\varphi=0\ \ \ \Leftrightarrow\ \ r=\underset{(0)}{\sigma}\pm\sqrt{\underset{(0)}{\sigma}^2-\underset{(0)}{\omega}^2}\ ,\end{equation}
and this requires 
\begin{equation}\label{kp57a}
\underset{(0)}{\sigma}^2\geq\underset{(0)}{\omega}^2\ .
\end{equation}
Always assuming that $A(u)\neq0$ we see from this inequality that \emph{provided $\theta\neq\frac{\pi}{2}$ and $\underset{(0)}{S}\neq0$ we must have}
\begin{equation}\label{kp58a}
1\leq\frac{1}{4}A^2\underset{(0)}{S}^2\sin^4\theta<\frac{1}{4}A^2\underset{(0)}{S}^2\ .
\end{equation}
But this condition by itself excludes Kinnersley's model as a special case. Consequently we can say that the Kerr analogue of Kinnersely's field of an arbitrarily accelerating point mass has a curvature singularity on a ring of radius $|\underset{(0)}{S}|$ undergoing rectilinear motion with arbitrary acceleration.

From the foregoing we see that asymptotically the matter distribution created by the accelerating, spinning source is qualitatively similar to that of Kinnersley's accelerating point mass except that the propagation direction of the light--like matter in the spinning case has shear on account of the acceleration/spin interaction. A further manifestation of the asymptotic similarity between the spinning case and Kinnersley's non--spinning case is displayed by comparing the formulas (\ref{kp4}) and (\ref{kp25'}) for large values of $r$. We begin by writing (\ref{kp25'}) as
\begin{equation}\label{kp39a}
X^i=w^i(u)+r\,K^i\ \ \ {\rm with}\ \ \ K^i=k^i+\frac{1}{r}\,U^i\ .
\end{equation}
Written out explicitly
\begin{equation}\label{kp40a}
U^i=P_0^2\left (\frac{\partial k^i}{\partial\xi}\frac{\partial k^j}{\partial\eta}-\frac{\partial k^j}{\partial\xi}\frac{\partial k^i}{\partial\eta}\right )s_j\ .
\end{equation}
With $k^i$ and $P_0$ given by (\ref{kp5}) and (\ref{kp6}) we can write
\begin{equation}\label{kp41a}
P_0^2\left (\frac{\partial k^i}{\partial\xi}\frac{\partial k^j}{\partial\eta}-\frac{\partial k^j}{\partial\xi}\frac{\partial k^i}{\partial\eta}\right )=\epsilon_{ijkl}\ v^k\,k^l\ ,
\end{equation}
where $\epsilon_{ijkl}$ is the four dimensional Levi Civita permutation symbol (we take $\epsilon_{1234}=+1$). If we define the spin tensor 
\begin{equation}\label{kp42a}
s_{ij}=\epsilon_{ijkl}\,s^k\,v^l=-s_{ji}\ \ \ \Leftrightarrow\ \ \ s_i=\frac{1}{2}\epsilon_{ijkl}\,s^{jk}\,v^l\ ,
\end{equation}
then
\begin{equation}\label{kp43a}
U^i=s^i{}_j\,k^j\ \ \ \Rightarrow\ \ \ K^i=\left (\delta^i_j+\frac{1}{r}\,s^i{}_j\right )k^j\ ,
\end{equation}
demonstrating that asymptotically $K^i$ only differs from $k^i$ by an infinitesimal Lorentz transformation. For the axially symmetric case
\begin{equation}\label{kp44a}
s_{ij}=\underset{(0)}{S}\,\epsilon_{ij34}\ ,
\end{equation}
and using (\ref{kp20})
\begin{eqnarray}
K_i\,dX^i&=&k_i\,dX^i+\frac{1}{r}\underset{(0)}{S}(k^2dX-k^1dY)\nonumber\\
&=&k_i\,dX^i+\frac{1}{r}\underset{(0)}{S}\sin\theta\,(\cos\phi\,dY-\sin\phi\,dX)\ , \nonumber\\\label{kp45a}
\end{eqnarray}
with $k_i\,dX^i$ given by (\ref{kp35}). Making use of (\ref{kp32}) and (\ref{kp39}) we find that
\begin{equation}\label{kp46a}
K_i\,dX^i=\lambda^{(3)}-\frac{1}{r}\underset{(0)}{S}\sin\theta\,\lambda^{(2)}=N_i\,dx^i\ ,\end{equation}
defining $N_i$ in coordinates $x^i=(\theta, \phi, r, u)$. In terms of the basis $f^i$, $e^i$, $l^i$, $n^i$ we therefore have
\begin{equation}\label{kp47a}
N^i=n^i-\frac{1}{r}\underset{(0)}{S}\sin\theta\,e^i\ ,\end{equation}
and thus it is immaterial whether we use $n^i$ or $N^i$ in the leading term in the energy--momentum--stress tensor components.

\section{Conclusions}\label{sec_conclusions}

We have shown how to construct an exact solution for the gravitational field of an arbitrarily accelerating and rotating point mass in General Relativity. This Kerr--like analogue of Kinnersely's field of an arbitrarily accelerating point mass has a curvature singularity on a ring of radius $|\underset{(0)}{S}|$ undergoing rectilinear motion with arbitrary acceleration. Asymptotically the matter distribution created by the accelerating, spinning source is qualitatively similar to that of Kinnersley's accelerating point mass except that the propagation direction of the light--like matter in the spinning case has shear on account of the acceleration/spin interaction. The uniqueness of the solution given here is a topic for further study and is related to the uniqueness of the Kerr solution.

\begin{acknowledgments}
This work was funded by the Deutsche Forschungsgemeinschaft (DFG, German Research Foundation) through the grant PU 461/1-2 -- project number 369402949 (D.P.). 
\end{acknowledgments}

\bibliographystyle{unsrtnat}
\bibliography{kinnersley_bibliography}
\end{document}